\documentclass[aps,prd,twocolumn,groupedaddress]{revtex4}

\usepackage{amsmath,amssymb}
\usepackage{graphicx}

\begin{document}

\title{Quasinormal mode expansion and the exact solution\\ of the Cauchy problem for wave equations}

\author{Nikodem Szpak}
\affiliation{Institute for Theoretical Physics, J.W.Goethe University Frankturt, Germany}

\date{\today}

\begin{abstract}
Solutions for a class of wave equations with effective potentials are obtained by a method of a Laplace-transform. Quasinormal modes appear naturally in the solutions only in a spatially truncated form;  their coefficients are uniquely determined by the initial data and are constant only in some region of spacetime --- in contrast to normal modes. This solves the problem of divergence of the usual expansion into spatially unbounded quasinormal modes and a contradiction with the causal propagation of signals. It also partially answers the question about the region of validity of the expansion. Results of numerical simulations are presented. They fully support the theoretical predictions.
\end{abstract}


\maketitle


\def\R{\mathbb{R}}
\def\C{\mathbb{C}}

\section{Introduction}

We study a class of wave equations \footnote{called also Klein-Gordon equations. Our nomenclature is motivated by the fact that the potential vanishes asymptotically and therefore the equation does not describe a ``massive'' field.} in the form
\begin{equation} \label{waveeq}
  \left[\frac{\partial^2}{\partial t^2} - \frac{\partial^2}{\partial x^2} + V(x) \right] \psi(t,x) = 0,
\end{equation}
describing propagation of a real wave function, $\psi: \R^+\times\R \rightarrow \R$ or $\C$, in presence of a real static potential $V(x)$ and satisfying initial conditions 
\begin{equation}
  \psi(0,x) = f(x),\qquad \dot{\psi}(0,x) = g(x).
\end{equation}
Equations of this type appear, among others, in the ana\-ly\-sis of linear perturbations around static solutions of nonlinear wave equations \cite{NS-Mth, NS-TMP}. In particular, they describe propagation of scalar, electromagnetic and gravitational perturbations of black holes which are static solutions of the Einstein equations (Regge-Wheeler eq. \cite{RegWhe}, Zerilli eq. \cite{ZerPert}). By a suitable change of variables and separation of the angular dependence, the original wave equations on a curved geometry can be brought to the above form (\ref{waveeq}), which contains only second partial derivatives and an effective potential $V$ \cite{Nol-rev}. The range of the new variable $x$ is determined by the coordinate change and usually extends from $-\infty$ (corresponds to the black hole horizon) to $+\infty$ (corresponds to $r\rightarrow\infty$). The potential contains all necessary information which is needed to describe scattering of the (partial) waves on a given geometry.

Study of the linear perturbations of black holes has led to the observation that solutions can be expressed as a sum over a countable set of modes, called \textit{quasinormal modes}  (to be distinguished from normal modes). This representation is universal in the sense that the set of modes depends only on the form of the potential and not on the initial data. Moreover, the modes possess different damping factors and hence only a few least damped modes are observed, what makes the picture simplier and even more universal. The parameters of a black hole are encoded in the set of modes and may be read off by future astrophysical observations of gravitational radiation (LIGO or VIRGO) coming from the vicinity of black holes. 

In the following we will leave the interpretation of waves travelling on a black-hole background and study the mathematical properties of the equation (\ref{waveeq}), since it has wider applications in various areas of mathematical physics.

Originally, quasinormal modes (QNMs) were defined as solutions of an eigenvalue problem resulting from time separation in the wave equation (\ref{waveeq}), after imposition of (artificial) boundary conditions of an outgoing wave at both ``ends'': $x\rightarrow\pm\infty$ \cite{Leaver}. It has been shown for several potentials that there exists an infinite countable set of such modes, what made them a good candidate for representation of solutions. Later, Sun and Price \cite{SunPrice} and Nollert and Schmidt \cite{NolSch} used a Laplace transform approach and a Green's function technique to show how a sum over quasinormal modes appears in the solution. The boundary conditions of outgoing waves, which seemed not to be uniquely defined (for $Re(s)<0$), have been replaced by a more familiar one of square integrability (for $Re(s)>0$) and analytic continuation (to $Re(s)<0$), and have been reproduced in the calculation in a uniquely defined way. Although the technique has been successfully applied for the general definition of QNMs and for indication how they appear in the solution, it has never been studied (to our best knowledge) more rigorously in a general setting, in particular, to clarify in which region of spacetime the quasinormal modes describe an exact solution and what is the form of the solution outside the region. There exists only literature concerning special potentials \cite{Beyer, NolPri, PriHus, LeungCavity}. Moreover, dealing with QNMs leads to several mathematical problems, which require better understanding of the the role they play in the solution.

First of them is that the profiles of QNMs are spatially unbounded. It makes every series built of them divergent at big distances. Although, the time dependence of the quasinormal modes is different than by normal modes -- they oscillate and decay, what makes the divergence more harmless for late times. Yet the problem of early and intermediate times remains open. It causes also difficulties in the definition of excitation coefficients -- usual scalar product-like formulas used for initial data and the modes lead to divergencies \cite{SunPrice, And-Excit}.

Second problem is completeness of the sum of quasinormal modes, i.e. if every solution of the wave equation can be (uniquely) represented by QNMs. It turns out that it essentially depends on the analytic and asymptotic properties of the potential, e.g. long range potentials usually make QNMs incomplete (by causing a tail due to a scattering at large distances \cite{Hod, ChingTails}) and discontinuities in the potential improve completeness (by increase in the number of modes \cite{NollertSteps, ChingCompl}).

In this Article we address the first problem and not the second. In particular, we show that the divergencies have their origin in too careless use of analogies with normal modes, thus leading to mathematical inconsistencies. In the evolution of smooth initial data, in presence of a well behaving potential, divergencies are not expected. We perform more careful calculations and obtain a well-defined representation of solutions, which in some region of spacetime (roughly, common future light cone of all initial data) takes the form of a sum over \textit{spatially truncated quasinormal modes} with constant coefficients. The coefficients can be explicitly calculated from the initial data by a scalar product-like formula involving initial data and QNMs. Though in general, they may depend on time and space, but it turns out that they become constant in a special region of spacetime, where the quasinormal modes become meaningful. The appearance of spatially truncated profiles of QNMs solves the problem of divergence at large distances and restores the principle of a causal propagation. We argue that this is the correct way to unterstand the quasinormal modes (cf. \cite{Nol-rev}) since the result is a natural consequence of solving the wave equation (\ref{waveeq}) via Laplace transform and treating the asymptotic behaviour of the integrand in the inverse transform with appropriate care, with no other ad hoc assumptions about the form of the result. 

The problem of completeness \cite{ChingCompl, PriHus} will appear in the calculations. We will introduce necessary assumptions to eliminate it. As will be more transparent during the calculations, these two problems are quite unrelated and therefore we find it justified to study them separately. 

Considerations presented in this Article are not fully rigorous. We are still working on a corresponding theorem, which shall contain analysis of the class of evolved initial data (domains of operators) and conditions on the analytic and asymptotic properties of the potential. Therefore we would characterize these results as \textit{one step more rigorous} than other general results known so far. 

This paper is organized as follows. In the second section we revise the theory of normal modes in order to understand (later) some essentiall differences between normal and quasinormal ones which might easily be overlooked in a less rigorous approach. In the third section we present a general theory of quasinormal modes and emphasize restrictions in its validity. In the fourth section we present our main result -- calculation of the Green's function, which contains construction of its asymptotic form for big $s$ (Laplace transform parameter) and appearance of quasinormal modes with constant coefficients in some region of a spacetime. In the fifth section we briefly discuss a connection between the region of validity of the quasinormal mode expansion, its completeness and properties of the potential. In the sixth section we present an example, which can be solved analytically, with results of a numerical simulation, which support our theoretical picture.

\section{Normal modes}

The simplest proofs of normal mode expansions rely on the properties of the corresponding operator $-\partial_x^2 + V(x)$ defined together with boundary conditions on a given space of functions. If the operator is self-adjoint the spectral decomposition into its eigenfunctions leads immediately to the normal mode expansion. Since in the case of quasinormal modes the corresponding operator is no more self-adjoined (different boundary conditions) no simple eigenfunction expansion exists and hence one cannot take much advantage of the `algebraic' methods. This is the reason why we present an `analytic' proof of normal mode expansion. The same technique will be later used for quasinormal modes.

\subsection{$V\equiv 0$}

\def\ra{\rightarrow}
\def\L{\mathcal{L}}
\def\hpsi{\hat{\psi}}
\def\hphi{\hat{\phi}}
\def\hG{\hat{G}}

Consider the wave equation in one dimension on an interval $[a,b]$
\begin{equation} \label{waveeq1d}
  \ddot\psi(t,x) - \psi''(t,x) = 0.
\end{equation}
with Sturm-Liouville type boundary conditions at $x=a$ and $x=b$ and initial conditions: $\psi(0,x) = f(x)$, $\dot{\psi}(0,x) = g(x)$. It is well known that the solution can be expressed as a sum over normal modes
\begin{equation}
  \psi(t,x) = \sum_n a_n\; \phi_n(x)\; e^{i \omega_n t},
\end{equation}
where $\phi_n(x)$ are normal modes, solving a corresponding eigenvalue problem (obtained from (\ref{waveeq1d}) by time separation) with an eigenvalue $\omega_n$, which plays a role of a (real) frequency of the mode. $a_n$ are excitation coefficients, uniquely determined by the initial data $f(x), g(x)$. Although the result is standard, we want to derive it here, in order to be able to compare it to the derivation of a quasinormal mode expansion and understand the differences. That is why the method chosen here is the same as the one introduced by Nollert and Schmidt \cite{NolSch} for definition of quasinormal modes.

Starting from (\ref{waveeq1d}) we perform a Laplace transform of the wave function
\begin{equation}
  \hpsi(s,x) = \L [\psi(\cdot,x)](s) = \int_0^\infty e^{-st}\; \psi(t,x)\; dt 
\end{equation}
and obtain an inhomogeneous ODE
\begin{equation} \label{weq-inh}
  s^2 \hpsi(s,x) - \hpsi''(s,x)  = s f(x) + g(x) \equiv j(s,x)
\end{equation}
This equation can be solved by a Green's function technique
\begin{equation} \label{ODE-hG}
  \left[ s^2 - \partial_x^2 \right] \hG(s,x,x') = \delta(x-x'),  
\end{equation}
\begin{equation} \label{hpsi-hG}
  \hpsi(s,x) = \int_{a}^{b} \hG(s,x,x')\; j(s,x')\; dx'.  
\end{equation}
A proper Green's function has to be constructed, i.e. which, beside the differential equation (\ref{ODE-hG}), must also satisfy the same Sturm-Liouville boundary conditions as $\psi$ does. As is well known, too, the following construction leads to the required result \cite{ByrFul}. Take two solutions of a corresponding to (\ref{weq-inh}) homogeneous ODE (with $j(s,x)$ replaced by zero)
\begin{equation} \label{homODE}
  \left[ s^2 -\partial_x^2 \right] \hphi_\pm(s,x) = 0,  
\end{equation}
which satisfy one boundary condition each, $\hphi_-$ the left one (at $x=a$) and $\hphi_+$ the right one (at $x=b$).
Then the proper Green's function reads
\begin{equation}
  \hG(s,x,x') = -\frac{\hphi_-(s,x_<)\; \hphi_+(s,x_>)}{W[\hphi_-,\hphi_+](s)},
\end{equation}
with $x_<=\min(x,x')$ and $x_>=\max(x,x')$, and $W[\hphi_-(x),\hphi_+(x)](s) \equiv \hphi_-(x)\;\hphi'_+(x) - \hphi'_-(x)\;\hphi_+(x)$ a Wronski determinant, independent of $x$. 

\begin{figure}
\includegraphics[width=0.8\columnwidth]{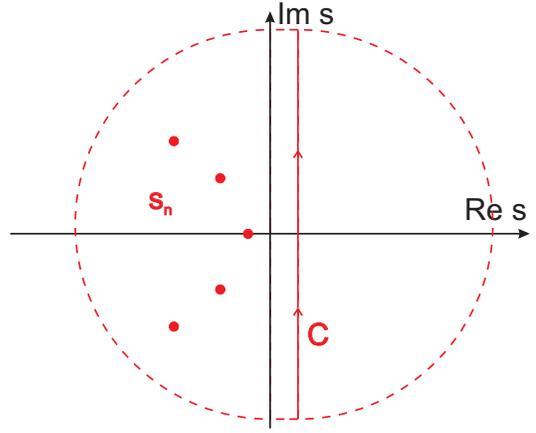}
\caption{Contour of integration ($C$) for the inverse Laplace transform.\label{fig_invL}}
\end{figure}

Now, we are ready to invert the Laplace transform
\begin{equation} \label{psi-invL}
  \psi(t,x) = \L^{-1}[\hpsi(\cdot,x)](t) 
            = \frac{1}{2\pi i}\int_C e^{st}\; \hpsi(s,x)\; ds,  
\end{equation}
where the integration in the complex plane is to be performed parallel to the imaginary axis, to the right of all poles (see Fig. \ref{fig_invL}). Making use of the relation (\ref{hpsi-hG}) and the definition of $j(s,x)$ we can express it, after some manipulations, as
\begin{equation} \label{psi-G}
\begin{split}
  \psi(t,x) &= \int_{a}^b \frac{\partial}{\partial t}\L^{-1}[\hG(\cdot,x,x')](t)\; f(x')\; dx' \\
            &+ \int_{a}^b \L^{-1}[\hG(\cdot,x,x')](t)\; g(x')\; dx',  
\end{split}              
\end{equation}
or shortly
\begin{equation}
\begin{split}
  \psi &= \frac{\partial}{\partial t}\L^{-1}[\hG] \ast f + \L^{-1}[\hG] \ast g \\
       &\equiv \frac{\partial}{\partial t} G \ast f + G \ast g.
\end{split}         
\end{equation}
Here, the initial data separate from the rest of the problem, i.e. the boundary conditions and the wave equation itself. The whole universal information about solutions of the problem is contained in the Green's function
\begin{equation} \label{G-invL}
  G(t,x,x') \equiv \L^{-1}[\hG(\cdot,x,x')](t) = \frac{1}{2\pi i}\int_C e^{st}\; \hG(s,x,x')\; ds  
\end{equation}
or, more explicitly,
\begin{equation} \label{G}
  G(t,x,x') = \frac{-1}{2\pi i}\int_C e^{st}\; \frac{\hphi_-(s,x_<)\; \hphi_+(s,x_>)}{W[\hphi_-,\hphi_+](s)}\; ds.
\end{equation}
A theorem of Weyl \cite{Richt} says that the functions $\hphi_\pm(s,x)$ are entire functions of the parameter $s$. Therefore the function $\hG(s,x,x')$ and the integrand are meromorphic in the whole complex plane, with possible poles only at zeros of the Wronskian $W[\hphi_-,\hphi_+](s)$. 

The simplest way to calculate the integral is to close the contour by a semicircle at infinity, either at the left or right hand side of the complex plane (Fig. \ref{fig_invL}). The contribution of the integral along the semicircle depends, in general, on the values of $t,x$ and $x'$. Indeed, it can be seen by an example, where the boundary conditions $\psi(t,a)=\psi(t,b)=0$ have been chosen\footnote{The same calculation can be carried out for any Sturm-Liouville type boundary conditions.}, that
\begin{equation}
  \hG(s,x,x') = \frac{[e^{s(x_<-a)}-e^{s(a-x_<)}][e^{s(x_>-b)}-e^{s(b-x_>)}]}{2s[e^{s(a-b)}-e^{s(b-a)}]}
\end{equation}
(more details can be found in Appendix \ref{AppA}). Its asymptotic form for $|s|\gg 1$, $Re(s)<0$ is
\begin{equation} \label{hGNM-}
  \hG(s,x,x') \simeq - \frac{e^{s|x-x'|}}{2s}
\end{equation}
and for $|s|\gg 1$, $Re(s)>0$
\begin{equation} \label{hGNM+}
  \hG(s,x,x') \simeq \frac{e^{-s|x-x'|}}{2s}.
\end{equation}
It implies that the integrand in (\ref{G}) behaves asymptotically like
\begin{equation}
  -\frac{e^{s(t+|x-x'|)}}{2s},\;Re(s)<0\quad\mbox{and}\quad \frac{e^{s(t-|x-x'|)}}{2s},\;Re(s)>0,
\end{equation}
respectively. The first function, for $Re(s)<0$, is always small for $t\ge 0$ and the integral over the left semicircle vanishes for all $t>0$. The second function, for $Re(s)>0$, is small only for $t\le |x-x'|$ and the integral over the right semicircle vanishes only for $t<|x-x'|$.

Since, by definition of the inverse Laplace transform, all poles in (\ref{G}) lay to the left of the original contour $C$, closing the contour to the right is free from the contribution of poles. In contrary, closing the contour to the left produces an additional contribution of residua calculated at the positions of the poles. The residua correspond to the values of $s$ at which the Wronskian $W[\hphi_-,\hphi_+](s)$ vanishes. At these points both functions are proportional
\begin{equation}
  \hphi_+(s_n,x) = c_n \hphi_-(s_n,x) \equiv c_n \phi_n(x)
\end{equation}
with some proportionality constant $c_n$. The integral in (\ref{G}) over a closed (to the left) contour may, by the Cauchy theorem, be expressed as a sum over the residua 
\begin{equation}
\begin{split}
  &G(t,x,x') = \\
  &= \frac{-1}{2\pi i}\cdot 2\pi i \sum_n \mbox{Res}\left[e^{s t} \frac{\hphi_-(s,x_<)\; \hphi_+(s,x_>)}{W(s)}\right]_{s=s_n}\\
  &= - \sum_n e^{s_n t} \hphi_-(s_n,x_<)\; \hphi_+(s_n,x_>) \mbox{Res}\left[\frac{1}{W(s)}\right]_{s=s_n}\\
  &= - \sum_n e^{s_n t} c_n\; \phi_n(x_<)\; \phi_n(x_>) \frac{1}{W'(s_n)}.
\end{split}  
\end{equation}
Introducing ``normalization constants'' $\alpha_n \equiv -c_n/W'(s_n)$ we finally arrive at
\begin{equation} \label{Gmodes}
  G(t,x,x') = \sum_n \alpha_n e^{s_n t} \phi_n(x)\; \phi_n(x')
\end{equation}
for every $t>0$, while closing the contour to the right gives simple
\begin{equation} \label{Gzero}
  G(t,x,x') = 0
\end{equation}
for $t<|x-x'|$. This is the main result for normal mode expansion. 
It is important to note, that for $t<|x-x'|$ the contour me be closed to either side, giving either a sum over modes or zero in the result, what leads to the conclusion that the sum over the modes converges in this case to zero. Combining both results into one formula gives
\begin{equation} \label{Gnm-theta}
  G(t,x,x') = \theta(t-|x-x'|)\cdot\sum_n \alpha_n e^{s_n t} \phi_n(x)\; \phi_n(x'),
\end{equation}
making the causality property explicit.

For completeness, let's calculate, in the last step, the wave function, using (\ref{psi-G})
\begin{equation}
\begin{split}
  \psi(t,x) &= \int_a^b \sum_n \alpha_n s_n e^{s_n t} \phi_n(x)\; \phi_n(x')\; f(x')\; dx'\\
            &+ \int_a^b \sum_n \alpha_n e^{s_n t} \phi_n(x)\; \phi_n(x')\; g(x')\; dx'.
\end{split}           
\end{equation}
Introducing the excitation coefficients
\begin{equation}
  A_n = \alpha_n \int_a^b \left[ s_n\; f(x') + g(x') \right]\; \phi_n(x')\; dx'
\end{equation}
the solution takes a compact form
\begin{equation}
  \psi(t,x) = \sum_n A_n\; \phi_n(x)\; e^{s_n t}.
\end{equation}

This result shows that (i) the form of the modes depends only on the form of the wave equation and boundary conditions and not on the initial data, (ii) the excitation coefficients are uniquely determined from the initial data, (iii) the expansion in a sum over modes is possible for every time $t>0$, and (iv) the causality condition holds, i.e. initial data may influence the solution only within their future light cone (cf. (\ref{Gnm-theta})).

\subsection{$V\neq 0$}

An analogous construction can be carried out if the waves propagate in presence of a potential, like in (\ref{waveeq}),
as long as the boundary conditions are of the Sturm-Liouville type and are imposed at the ends of a finite interval.

Let the boundary conditions have the general form:
\begin{eqnarray}
  \beta\; \psi(t,a)-\alpha\; \frac{\partial \psi}{\partial x}(t,a)&=&0\\
  \delta\;\psi(t,b)-\gamma\; \frac{\partial \psi}{\partial x}(t,b)&=&0.
\end{eqnarray}
with $\alpha, \beta,\gamma, \delta \in \R$. By the technique described in more detail in section \ref{SecGreensF} it can be shown that
\begin{multline}
 \hphi_-(s,x) =\\ \frac{1}{2}\left[
      e^{s(x-a)}\left(\alpha+\frac{\beta}{s}\right) + e^{-s(x-a)}\left(\alpha-\frac{\beta}{s}\right)\right]
      \left(1+\mathcal{O}(s^{-1})\right) \nonumber
\end{multline}       
\begin{multline}
 \hphi_+(s,x) =\\ \frac{1}{2}\left[
      e^{s(x-b)}\left(\gamma+\frac{\delta}{s}\right) + e^{-s(x-b)}\left(\gamma-\frac{\delta}{s}\right)\right]
      \left(1+\mathcal{O}(s^{-1})\right).
\end{multline}       

Then, first the approximate Wronskian $W[\hphi_-,\hphi_+](s)$ and next the approximate Green's function can be obtained with the same precision (see Appendix \ref{AppB} for more details). Considering its asymptotic behaviour for $|s|\gg 1$ one finds the same formulas as in the previous section ($V=0$), namely (\ref{hGNM-}) and (\ref{hGNM+}). Their consequence is the normal mode decomposition
\begin{equation} 
  G(t,x,x') = \sum_n \alpha_n e^{s_n t} \phi_n(x)\; \phi_n(x')
\end{equation}
for every $t>0$ and
\begin{equation}
  G(t,x,x') = 0
\end{equation}
for $t<|x-x'|$. For nonnegative potentials all $s_n$ must be purely imaginary (because $-s_n^2$, the eigenvalues of $-\partial_x^2+V(x)$, must be real and positive). For negative potentials $s_n$ must be either purely imaginary or real positive (in all cases $s_n^2\leq -\inf_{x\in[a,b]} V(x)$).

\subsection{Leaky cavity}

Even for some non-Sturm-Liouville boundary conditions, which lead to non-self-adjointness of $-\partial_x^2+V(x)$ on the corresponding domain, the mode expansion may be possible for all times $t>0$. Again the same technique of constructing approximate Green's function can be used for the proof.

Consider boundary conditions of the form
\begin{eqnarray}
  \psi(t,0)&=&0,\\
  \frac{\partial \psi}{\partial t}(t,L) + \gamma\; \frac{\partial \psi}{\partial x}(t,L)&=&0.
\end{eqnarray}

The left boundary $x=0$ is fully reflecting and the right one is leaky -- part of the incoming signal from the left will be reflected and part will be transmitted and lost from the domain $[a,b]$. The parameter $\gamma$ measures the leakage: for $\gamma=0$ the boundary is fully reflecting and for $\gamma=1$ it is transparent for waves coming from the left.

The boundary conditions for $\psi(t,s)$ translate into conditions on the Laplace transformed functions
\begin{eqnarray}
  \hphi_-(s,0)&=&0,\\
  s\; \hphi_+(s,L) + \gamma\; \frac{\partial \hphi_+}{\partial x}(s,L)&=&0 \label{LCbc2}
\end{eqnarray}
(we have assumed $\psi(0,L)=0$ for simplicity, otherwise this term should appear on the r.h.s. of (\ref{LCbc2})).

The solutions are
\begin{equation}
 \hphi_-(s,x) = e^{sx} - e^{-sx}
\end{equation}
\begin{multline}
 \hphi_+(s,x) =\\ \frac{1}{2}\left[
      e^{s(x-L)}\left(\gamma-1\right) + e^{-s(x-L)}\left(\gamma+1\right)\right]. \nonumber
\end{multline}       
Next, the Wronskian $W[\hphi_-,\hphi_+](s)$ and finally the Green's function can be constructed (see Appendix \ref{AppC} for more details). The asymptotic behaviour for $|s|\gg 1$ is the same as for normal modes discussed above and given by (\ref{hGNM-}) and (\ref{hGNM+}). Analogously the mode decomposition follows
\begin{equation}
  G(t,x,x') = \sum_n \alpha_n e^{s_n t} \phi_n(x)\; \phi_n(x')
\end{equation}
for every $t>0$ and
\begin{equation}
  G(t,x,x') = 0
\end{equation}
for $t<|x-x'|$. The essential difference is that the frequencies $s_n$ are no longer purely imaginary or real. They can be complex and for nonnegative potentials must lay in the left half of the complex plane ($Re(s_n)<0$). In case of positive potentials there may be additional $s_n$ which lay on the positive half of the real axis ($s_n\in\R^+$).

\section{Quasinormal modes}

Consider now the same wave equation with a potential
\begin{equation}
  \left[\frac{\partial^2}{\partial t^2} - \frac{\partial^2}{\partial x^2} + V(x) \right] \psi(t,x) = 0,
\end{equation}
but on an infinite axis, $x\in\R$, and initial conditions
\begin{equation}
  \psi(0,x) = f(x),\qquad \dot{\psi}(0,x) = g(x).
\end{equation}
About the potential we assume that it is real, continuous and vanishes faster than $1/|x|$ as $x\ra\pm\infty$.
To construct a similar set of modes a kind of a boundary condition at $x\ra\infty$ is needed. Mathematically it corresponds to the definition of a domain of solutions. Historically, one has used a boundary condition of an asymptotically outgoing wave. This bases on the experience from the scattering theory. Although it is to some extent correct, it proves unsatisfactory in more detailed calculations, namely, for $Re(s)<0$ this boundary condition fails to be unique. Nollert and Schmidt \cite{NolSch} proposed a procedure, based on a Laplace transform and analytic properties of the transformed Green's function, which leads to uniquely defined quasinormal modes. We have demonstrated the method of a Laplace transform in the previous section. What remains, is to see how the change of the boundary conditions (and therefore the domain of solutions) causes the quasinormal modes to replace the normal ones.

Performing again the Laplace transform, which is defined for $Re(s)>0$,
\begin{equation}
  \hpsi(s,x) = \L [\psi(\cdot,x)](s) = \int_0^\infty e^{-st}\; \psi(t,x)\; dt, 
\end{equation}
we obtain the inhomogeneous ODE
\begin{equation} 
  \left[- \partial_x^2 + s^2 + V(x)\right]\hpsi(s,x) = s f(x) + g(x) \equiv j(s,x).
\end{equation}
We solve it by a Green's function technique
\begin{equation}
  \left[ - \partial_x^2 + s^2 + V(x) \right] \hG(s,x,x') = \delta(x-x'),  
\end{equation}
\begin{equation} \label{hpsi-hG1}
  \hpsi(s,x) = \int_{a}^{b} \hG(s,x,x')\; j(s,x')\; dx'.  
\end{equation}
The proper Green's function, again, is constructed from two solutions of the corresponding homogeneous ODE
\begin{equation} \label{homODE1}
  \left[ -\partial_x^2 + s^2 + V(x) \right] \hphi_\pm(s,x) = 0,  
\end{equation}
which have the property, that $\hphi_\pm(s,x)$ satisfies the boundary condition as $x\ra\pm\infty$, respectively. The boundary condition is chosen to be the square integrability at infinity, i.e.
\begin{equation}
  \int_a^{\pm\infty} |\hphi_\pm(s,x)|^2\; dx < \infty
\end{equation}
for every $a\in\R$. As long as $Re(s)>0$, it is equivalent to the outgoing wave condition. It can be seen by an explicit construction of the asymptotic behaviour of the solution which is square integrable at the given end. Since the potential vanishes at infinity, we find
\begin{eqnarray}
  \hphi_-(s,x)\sim e^{+sx}\quad &\mathrm{as}&\quad x\ra -\infty\\
  \hphi_+(s,x)\sim e^{-sx}\quad &\mathrm{as}&\quad x\ra +\infty.
\end{eqnarray}
The inverse Laplace transform restores the time dependence, roughly, in the form of a multiplication by $\exp(st)$ (in the case of a single mode, at least), what gives
\begin{eqnarray}
  \phi_-(t,x)\sim e^{s(t+x)}\quad &\mathrm{as}&\quad x\ra -\infty\\
  \phi_+(t,x)\sim e^{s(t-x)}\quad &\mathrm{as}&\quad x\ra +\infty.
\end{eqnarray}
Both are purely outgoing waves, what agrees with the original idea of the quasinormal modes. 

A comment on a difference appearing in (\ref{homODE1}) relative to the previous section is in place here. Although the equation is nearly identical with (\ref{homODE}), the differential operator $-\partial_x^2 + V(x)$ is no more self-adjoint on the domain defined by the outgoing wave boundary conditions. Therefore its eigenvalues $-s^2$ do not have to be real. For normal modes it holds $s=i\omega$, $\omega\in\R$, what makes them purely oscillatory in time with the frequency $\omega$. This behaviour is not expected by quasinormal modes.

By these boundary conditions, the Green's function
\begin{equation}
  \hG(s,x,x') = -\frac{\hphi_-(s,x_<)\; \hphi_+(s,x_>)}{W[\hphi_-,\hphi_+](s)},
\end{equation}
with $x_<=\min(x,x')$ and $x_>=\max(x,x')$, is square integrable in both variables $x$ and $x'$. So is the transformed wave function $\hpsi(s,x)$ calculated from (\ref{hpsi-hG1}) if $j(s,x)$ (or equivalently $f(x)$ and $g(x)$) is of compact support.

The inversion of the Laplace transform goes the same way as in the preceding section, following exactly the formulas (\ref{psi-invL})-(\ref{G-invL}), and leads to the integral
\begin{equation} \label{G1}
  G(t,x,x') = \frac{-1}{2\pi i}\int_C e^{st}\; \frac{\hphi_-(s,x_<)\; \hphi_+(s,x_>)}{W[\hphi_-,\hphi_+](s)}\; ds.
\end{equation}
Here, the same idea of closing the integration contour by a semicircle at infinity seems most natural, but in this case the dependence of $\hphi_\pm(s,x)$ on $s$ is much less trivial than it was by normal modes. First, there is no theorem saying that they are entire functions of $s$. The special boundary conditions make them more complicated functions and poles as well as essential singularities producing branch cuts may occur. Fortunately, the poles in $\hphi_\pm(s,x)$ cancel with those in the Wronskian $W[\hphi_-,\hphi_+](s)$, but essential singularities and cuts in general survive. 

Second, it is not easy to determine the asymptotics of the integrand in (\ref{G1}) at big $|s|$. Its control is essential for estimation of the contribution of the integral along the semicircle at infinity. Let's first shortly recall what is popularly written in the literature: \cite{Nol-rev} reviews \textit{the state of the art} in the theory of quasinormal modes. It contains discussion about the divergence problem and speculations about time-dependence of the excitation coefficients (proposed by N. Andersson \cite{And-test}). Though, it does not bring firm answers, since on the mathematical level the problem is not treated carefully enough.

It looks natural that at positive times $t>0$ the exponent $\exp(st)$ in (\ref{G1}) goes to zero for big $|s|$ with $Re(s)<0$, so if the rest of the integrand is somehow bounded (here is the point!) then the contour can be closed by a semicircle at infinity on the left hand side. An important point is that the integrand is defined, via the Laplace transform, only for $Re(s)>0$ and its domain of definition must be extended to the left half-plane $Re(s)<0$. It can be done in various ways, although e.g. imposing the condition of square integrability proves wrong and useless, and the one of the asymptotics $\hphi_\pm\sim\exp(\mp sx)$ at $x\ra\pm\infty$ seems to be mathematically not well-defined, because the other linearly independent asymptotic solution $\exp(\pm sx)$ is subdominant relative to it and cannot be eliminated\footnote{At least without enhancing the asymptotic analysis of solutions into higher orders or super-/hyperasymptotics.}. The only correct method is the analytic continuation of $\hphi_\pm(s,x)$ from the right to the left half of the complex plane of $s$. Only this allows to use the Cauchy theorem and replace the complex integral by a sum over residua. 

Now, assume for a moment that there are no cuts in the complex plane, and use the Cauchy theorem. We arrive at the sum over modes, like in the previous section,
\begin{equation}  \label{G1-QNM}
  G(t,x,x') = \sum_n \alpha_n e^{s_n t} \phi_n(x)\; \phi_n(x'),
\end{equation}
but now the functions $\phi_n(x)$ are quasinormal modes. They are also defined by $\hphi_\pm(s,x)$ calculated at the position of the poles $s=s_n$ in the left complex half-plane, but they do not satisfy the boundary conditions in their original form of square integrability. Since we have chosen analytic continuation, their asymptotic form as $x\ra\pm\infty$ for $Re(s)<0$ remains explicitly the same (analytically continued, too) as it is for $Re(s)>0$. Moreover, every mode satisfies both boundary conditions, what gives
\begin{equation}
  \phi_n(x\ra -\infty)\sim e^{s_n x}\quad\mbox{and}\quad\phi_n(x\ra +\infty)\sim e^{-s_n x}.
\end{equation}
Since $Re(s_n)<0$, the quasinormal modes are unbounded at both ends. Optically, it looks like the asymptotic boundary conditions, criticized above as non-unique, due to an uncontrollable admixture of the other, subdominant asymptotic solution. Here, however, thanks to the analytic continuation from the right half-plane, where the asymptotic conditions are unique, the construction is unique, too. The spatial unboundedness throws a shadow of a doubt on their applicability in the expansion of solutions of the wave equation. Before we come to that question, let's first observe that already
the assumption of boundedness of the ``rest'' of the integrand in (\ref{G1}) is wrong. What we can expect, is rather a boundedness by an exponent $\sim\exp(\pm sx)$. Therefore, the procedure of closing the contour on the left and assuming that the contribution of the semicircle integral vanishes, seems only reasonable for late times $t\gg|x-x'|$. 

To answer the question more exactly, for which values of $t,x,x'$ the contour can be closed left or right with finite or vanishing contribution from the corresponding semicircle integral, we have to analyze the asymptotics of the integrand at big $|s|$ in more details. 

\section{Construction of the Green's function} \label{SecGreensF}

An essential observation is that the question if the contour of integration in the inverse Laplace transform of the Green's function may be closed on the right or left depends only on the asymptotics of the Green's function for big $|s|$. This, further, depends on the asymptotic behaviour of $\hphi_\pm(s,x)$ for big $|s|$, which satisfy the ordinary differential equation
\begin{equation} \label{ODE-LG}
  \partial_x^2 \hphi(s,x) = \left[ s^2 + V(x) \right] \hphi(s,x).
\end{equation}
By means of the Liouville-Green approximation \cite{Olver} we are able to construct the asymptotic solutions of this equation to the leading order in $|s|$ and control the error. The same will be possible, in the next step, for the Green's function.

The theory says that there are two, so called, dominant solutions
\begin{eqnarray} \label{varphi1}
  \varphi_1(s,x) &=& u^{-1/4}\cdot e^{+\int u^{1/2}\; dx}\cdot (1 + \epsilon_1(s,x) )\\
  \varphi_2(s,x) &=& u^{-1/4}\cdot e^{-\int u^{1/2}\; dx}\cdot (1 + \epsilon_2(s,x) ) \label{varphi2}
\end{eqnarray}
with the bounds on the error functions $\epsilon_1, \epsilon_2$ and their derivatives
\begin{eqnarray} \label{eps1}
 |\epsilon_1(s,x)|,\;|u^{-1/2} \partial_x \epsilon_1(s,x)| &\leq& e^{\Upsilon_{a_1,x}} - 1 \\
 |\epsilon_2(s,x)|,\;|u^{-1/2} \partial_x \epsilon_2(s,x)| &\leq& e^{\Upsilon_{a_2,x}} - 1 \label{eps2}
\end{eqnarray}
with $\Upsilon_{a,b}$ defined (for our purpose) by
\begin{equation}
  \Upsilon_{a,b} = \left|\int_a^b |F'(x)|\; dx \right|
\end{equation}
and
\begin{equation}
  F(x) = \int \left[u^{-1/4}(u^{-1/4})''-vu^{-1/2} \right] dx.
\end{equation}
The auxiliary continuous functions $u$ and $v$ can be chosen freely and in our case must satisfy the condition $u(x)+v(x) = s^2 + V(x)$. The boundaries are $a_1=-\infty$ and $a_2=\infty$. The function $Re\int u^{1/2}\; dx$ must be nondecreasing on the intervals $(a_1,x)$ and $(x,a_2)$. $\Upsilon_{a_i,x}$ must be finite for every $x$ ($i=1,2$).

In our problem the most convenient choice is:
\begin{equation}
  u(x) = s^2,\qquad v(x) = V(x),
\end{equation}
with the choice of the complex root $u^{1/2} = s$. Then the solutions take the form
\begin{eqnarray}
  \varphi_1(s,x) &=& \frac{1}{\sqrt{s}}\; e^{+ sx}\cdot (1 + \epsilon_1(s,x) )\\
  \varphi_2(s,x) &=& \frac{1}{\sqrt{s}}\; e^{- sx}\cdot (1 + \epsilon_2(s,x) ).
\end{eqnarray}
The function $Re\int u^{1/2}\; dx = Re(s)x$ is obviously nondecreasing as $x$ increases on $(-\infty,\infty)$ as long as $Re(s)>0$. The auxiliary functions for the error estimates take quite simple form, too,
\begin{equation}
  F'(x) = -\frac{V(x)}{s}
\end{equation}
\begin{eqnarray}
  \Upsilon_{a_1,x} &=& \Upsilon_{-\infty,x} = \frac{1}{|s|} \int_{-\infty}^x |V(x)|\; dx\\
  \Upsilon_{a_2,x} &=& \Upsilon_{+\infty,x} = \frac{1}{|s|} \int_x^{+\infty} |V(x)|\; dx,  
\end{eqnarray}
and the errors are bounded according to (\ref{eps1}), (\ref{eps2}). Here it becomes transparent, that this construction works only for integrable potentials $V\in L^1(\R)$.

It can be easily seen from the bounds on $\epsilon_1, \epsilon_2$, that the solutions $\varphi_1, \varphi_2$ behave asymptotically like
\begin{eqnarray}
 \varphi_1(s,x) \sim e^{sx}\quad&\mathrm{as}&\quad x\ra -\infty \\
 \varphi_2(s,x) \sim e^{-sx}\quad&\mathrm{as}&\quad x\ra +\infty
\end{eqnarray}
and therefore satisfy the boundary conditions. Hence, we choose
\begin{equation}
  \hphi_-(s,x) = \varphi_1(s,x)\qquad \hphi_+(s,x) = \varphi_2(s,x).
\end{equation}
Now, the behaviour of the solutions $\hphi_\pm$ at big $|s|$ may be found. Since the potential $V(x)$ is continuous and vanishes asymptotically like $|V(x)| < C\; |x|^{-1-\varepsilon}$ with some $C, \varepsilon > 0$ for $|x|\ra\infty$, the integral $||V||\equiv\int_{-\infty}^{+\infty} |V(x)|\; dx$ is finite. We obtain the following estimates
\begin{equation}
  \Upsilon_{-\infty,x}\leq\frac{||V||}{|s|},\qquad \Upsilon_{+\infty,x}\leq\frac{||V||}{|s|},
\end{equation}
and the bounds on the errors 
\begin{eqnarray}
 |\epsilon_1(s,x)|,\;|u^{-1/2} \partial_x \epsilon_1(s,x)| &\leq& e^{||V||/|s|} - 1 \\
 |\epsilon_2(s,x)|,\;|u^{-1/2} \partial_x \epsilon_2(s,x)| &\leq& e^{||V||/|s|} - 1. 
\end{eqnarray}
These can be expanded for $|s|\gg 1$ to give
\begin{eqnarray}
 |\epsilon_1(s,x)|,\;|u^{-1/2} \partial_x \epsilon_1(s,x)| &=& \mathcal{O}\left(\frac{||V||}{|s|}\right) \\
 |\epsilon_2(s,x)|,\;|u^{-1/2} \partial_x \epsilon_2(s,x)| &=& \mathcal{O}\left(\frac{||V||}{|s|}\right).
\end{eqnarray}
Therefore
\begin{eqnarray}
  \hphi_-(s,x) &=& \frac{1}{\sqrt{s}}\; e^{+ sx}\cdot \left[1 + \mathcal{O}\left(\frac{||V||}{|s|}\right) \right]\\
  \hphi_+(s,x) &=& \frac{1}{\sqrt{s}}\; e^{- sx}\cdot \left[1 + \mathcal{O}\left(\frac{||V||}{|s|}\right) \right].
\end{eqnarray}
Further, it can be shown, that the error of their Wronskian is bounded in a similar way
\begin{equation}
  W[\hphi_-,\hphi_+](s) = -2\cdot\left[1 + \mathcal{O}\left(\frac{||V||}{|s|}\right) \right].
\end{equation}
Finally, the asymptotic form of the Green's function for big $|s|$ may be written
\begin{equation}
  \hG(s,x,x') = \frac{e^{-s|x-x'|}}{2s}\cdot\left[1 + \mathcal{O}\left(\frac{||V||}{|s|}\right) \right].
\end{equation}
It is important to note, that it asymptotically, for $|s|\ra\infty$, coincides with the free Green's function, i.e. for a wave equation with $V(x)\equiv 0$. It is not unexpected, since in the r.h.s of (\ref{ODE-LG}), in $s^2+V(x)$, the potential $V(x)$ is small compared to $s^2$ as $|s|\ra \infty$ (more exactly $s^2\gg ||V||$).

Unfortunately, this approximation is valid only for $Re(s)>0$, as already noted. The choice of $Re(s)<0$ in the construction of the dominant solutions leads to an exchange of the formulas (\ref{varphi1}), (\ref{varphi2}) for $\varphi_1, \varphi_2$. The reason is that the function $Re\int u^{1/2}\;dx$ must be nondecreasing in $x$, so it requires a change of $u^{1/2}$ from $s$ to $-s$. The whole construction can be repeated with the exchanged formulas, but then the functions $\hphi_\pm$ are constructed separately in the left and right complex half-plane of $s$. Obviously, they are not analytic in $s$, or more precisely, the functions $\hphi_\pm$ constructed for $Re(s)<0$ are not the analytic continuations of those constructed for $Re(s)>0$, what is required.

Having this in mind, let's come back to the analysis of the inverse Laplace transform for the Green's function (\ref{G1}). The integrand equals approximately, for big $|s|$, 
\begin{equation}
  \frac{e^{s(t-|x-x'|)}}{2s}\cdot\left[1 + \mathcal{O}\left(\frac{||V||}{|s|}\right) \right],
\end{equation}
what means, that the contour of integration may be closed on the right for $t-|x-x'|<0$. That leads to the conclusion
\begin{equation}
  G(t,x,x') = 0\quad \mathrm{for}\quad 0<t<|x-x'|.
\end{equation}
It is nothing else than the causality condition. The same result is obtained immediately in the free case ($V(x)\equiv 0$). It is not surprising that the presence of a potential in the wave equation does not change the causality of the wave propagation.

\begin{figure}
\includegraphics[width=0.8\columnwidth]{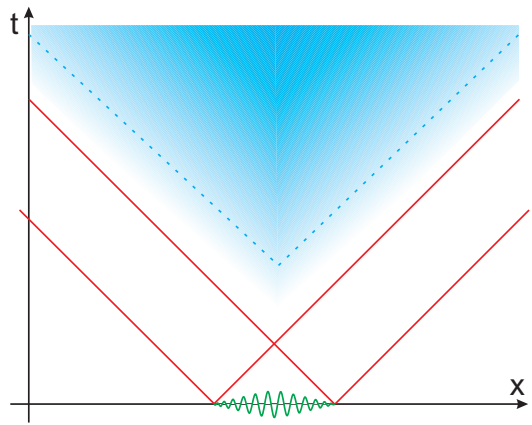}
\caption{Validity region (shaded) for the quasinormal mode expansion.\label{fig_valid}}
\end{figure}

Until now, we have reproduced ``one half'' of the result holding for normal modes (\ref{Gzero}), namely the causality condition for the Green's function. The ``second half'' (eq. (\ref{Gmodes}) for normal modes) should be the expansion of the Green's function into quasinormal modes. Unfortunately, the situation here is much more difficult: not only an expansion for any $t>0$ cannot be shown, because the integral along the semicircle at infinity on the left hand side may diverge, but also the expansion for $t>|x-x'|$, where the Green's function is nonzero, cannot be shown always. It can be proved only under further assumptions on the potential $V$. In general, it is expected that the expansion holds for some $t>t_0(x,x')>|x-x'|$. In \cite{Beyer} the Poeschl-Teller potential has been studied in detail and the function $t_0(x,x')$, which gives the earliest time of validity of the expansion, has been found. This region $(t,x)$ in evolution of initial data with compact support has the form like shown on the Fig. \ref{fig_valid}.

We want briefly sketch the technical difficulties which arise. The functions $\hphi_\pm(s,x)$, which for a given $s$ solve the ordinary differential equation (\ref{ODE-LG}) with one boundary condition, at $x\ra\pm\infty$, respectively, have a simple asymptotic behaviour on that end
\begin{eqnarray}
 \hphi_-(s,x) &\sim& e^{sx},\quad x\ra-\infty \\
 \hphi_+(s,x) &\sim& e^{-sx},\quad x\ra+\infty,
\end{eqnarray}
but a complicated behaviour on the other end
\begin{eqnarray}
 \hphi_-(s,x) &\sim& A(s) e^{sx} + B(s) e^{-sx},\quad x\ra+\infty \\
 \hphi_+(s,x) &\sim& C(s) e^{-sx}+ D(s) e^{sx},\quad x\ra-\infty.
\end{eqnarray}
The ``transmission'' coefficients $A(s)$ and $C(s)$ tend to $1$ and the ``reflection'' coefficients $B(s)$ and $D(s)$ tend to zero as $|s|\ra\infty$, but the rate of their convergence ($\sim s^{-n}$) is usually too slow to make the corresponding terms appearing in the Green's function, like
\begin{equation}
  B(s) C(s) e^{-s(x+x')}\quad \mathrm{or} \quad A(s) D(s) e^{s(x+x')},
\end{equation}
to vanish as $|s|\ra\infty$. This corresponds to the scattering of the waves on the potential, even in the limit of high frequencies. It causes that the contour of integration cannot be closed on the left hand side immediately after $t$ passes $|x-x'|$, what is the case for a free wave equation ($\exp[s(t-|x-x'|)]\ra 0$), but only later, when other contributions to the Green's function ($\sim\exp[s(t-x-x')]$ or $\exp[s(t+x+x')]$) vanish, too. Finally, the contour may be closed, and by the same, the expansion is valid, for times $t>t_0(x,x')\geq |x-x'|$ (Fig. \ref{fig_valid}). The function $t_0(x,x')$ is in general a complicated function of the positions $x,x'$ and depends essentially on the shape of the potential $V(x)$.

\begin{figure}
\includegraphics[width=0.8\columnwidth]{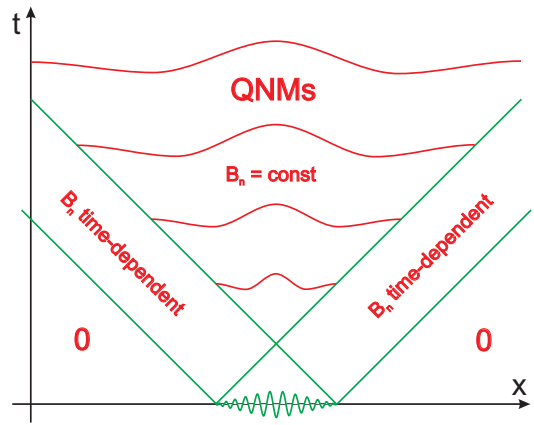}
\caption{Appearance of truncated quasinormal modes in evolution of initial data with compact support.\label{fig_truncated}}
\end{figure}

Another conclusion can be drawn from this construction. If the contour can be closed on the left hand side and the expansion into quasinormal modes holds, the modes will only appear in a spatially truncated form. The reason for the truncation is the causality condition together with the assumption that the initial data have compact support. They can influence only the region of spacetime laying inside their future light cone. In the outer region the solution must remain zero. In an optimal case (which can be realized by some potentials), when the expansion holds for all $t>t_0(x,x')\equiv |x-x'|$, the Green's function can be written as
\begin{equation} \label{G2}
  G(t,x,x') = \theta(t-|x-x'|)\cdot\sum_n \alpha_n e^{s_n t} \phi_n(x)\; \phi_n(x').
\end{equation}
It is the same result as for the normal mode expansion, but here the $\theta-$function plays an additional role: it truncates the modes at finite times and avoids their divergence at big distances (Fig. \ref{fig_truncated}). Furthermore, it is not indifferent now if we write the $\theta-$function in front of the sum or not, as it was by normal modes, because the sum itself is divergent just there, where $\theta(t-|x-x'|)$ equals zero, i.e. for $t<|x-x'|$. The sum over normal modes, in contrary, converged to zero in that region. 

The solution of the wave equation can be written in the form
\begin{equation} \label{Psi2}
  \psi(t,x) = \sum_n B_n(t,x)\; \phi_n(x)\; e^{s_n t},
\end{equation}
which is similar to that for normal modes, but the excitation coefficients become in general time and space dependent and are defined by
\begin{equation} \label{Bn2}
\begin{split}
  B_n(t,x) &= \alpha_n \int_{x-t}^{x+t} \left[ s_n\; f(x') + g(x') \right]\; \phi_n(x')\; dx'\\
           &+ \alpha_n \left[f(x-t)\; \phi_n(x-t) + f(x+t)\; \phi_n(x+t)\right],
\end{split}  
\end{equation}
what results from the presence of the $\theta-$function in (\ref{G2}). The $B_n(t,x)$ coefficients are zero outside the future light cone of the initial data. They get constant at points in spacetime $(t,x)$, where all initial data lay inside the past light cone of $(t,x)$, i.e. when all initial data ``can be seen'' (Fig. \ref{fig_truncated}).

\subsection{Case $V\equiv 0$}

It should be noted that the general result (\ref{G2}) contains the case of a free wave equation with $V(x)\equiv 0$. In this situation the exact solutions of (\ref{ODE-LG}) are
\begin{eqnarray}
  \hphi_\pm(s,x) &=& e^{\mp sx}\\
  W[\hphi_-,\hphi_+](s) &=& -2s
\end{eqnarray}
(where we have multiplied both solutions by $\sqrt{s}$ for convenience, what changes the Wronskian accordingly but has no influence on the Green's function). Then the Laplace transformed Green's function equals exactly
\begin{equation}
  \hG(s,x,x') = \frac{e^{-s|x-x'|}}{2s}.
\end{equation}
The only pole is placed at $s_0=0$. The corresponding quasinormal mode is trivial
\begin{equation}
  \phi_0(x) = 1,
\end{equation}
but what makes the solution of the wave equation nontrivial is the $\theta(...)$ factor in front of the QNM sum
\begin{eqnarray}
  G(t,x,x') &=& \theta(t-|x-x'|)\cdot \alpha_0\; e^{s_0 t} \phi_0(x)\; \phi_0(x') \nonumber \\
            &=& \frac{1}{2}\; \theta(t-|x-x'|),
\end{eqnarray}
where $\alpha_0 = \mbox{Res}[1/W(s)]_{s=s_0} = 1/2$. It is the well known Green's function for a free wave equation, which cannot be reproduced by formula (\ref{G1-QNM}) valid only for long times $t\gg|x-x'|$.

\section{Validity of the expansion}

A discussion when the quasinormal mode decomposition is a full solution of the Cauchy problem is in place here.
Beside the requirement of compact support of the initial data, in question are conditions on the potential $V(x)$. In this construction it has been first assumed that $V$ must be continuous and integrable, what is not a strong restriction. Though, further it has been assumed that the analytic continuation of the Laplace transformed Greens's function $\hG(s,x,x')$ should have no complex branch cuts. It is a complicated indirect condition on the potential. We do not want to go into a detailed discussion of this problem in this short Article. Its role is rather to clarify some questions regarding QNM decomposition on a heuristic level. We are working on a result in form of a theorem where the problem of branch cuts will be addressed with adequate care.

However, some general conclusions can be drawn without much analytical work. It has been observed that smooth potentials with exponential fall-off at infinity produce no cuts in $\hG(s)$. On the contrary, potentials with power-law decay usually produce a complex cut emanating from an essential singularity at $s=0$. This relation can be explained by the result of Hod \cite{Hod} which relates the fall-off of the solution at late times with the asymptotic behaviour of the potential (see also \cite{ChingTails}). Roughly, initial data evolving in presence of exponential potentials decay exponentially (like in P{\"o}schl-Teller potential studied in \cite{Beyer}) and in power-law potentials decay with a power-law (what is well known for black-hole backgrounds -- see \cite{Leaver} for detailed analysis of the Schwarzschild case and \cite{KokSch} for a review of all other black-holes). Since a sum of QNMs decays exponentially in time while the practise shows that solutions with a branch cut contribution in $\hG(s)$ fall-off slower, like power-law, it becomes plausible that exponential potentials should not give rise to a cut while long-range potentials produce brach cuts necessarily.

The second open problem is that of the earliest time $t_0(x,x')\ge |x-x'|$ when the QNM expansion begins to be valid. Its dependence on the shape of the potential seems to be very complicated, depending probably strongly on the analytic properties (differentiability) of $V(x)$. It is related to the scattering of waves in the limit of very high frequencies, where hopefully some results from the scattering theory may be used. It is known that discontinuities in the potential or its derivatives increase the number of modes \cite{NollertSteps, ChingCompl}.

\section{Example: $V = V_0 e^{-2r}$}

In this section we present an example of a wave equation, which can be, to a large extent, solved analytically.
We choose
\begin{equation}
  V(r) = V_0\;e^{-2r}
\end{equation}
and modify a little the original setup of the problem. Let now the spatial coordinate be defined on $r\in(0,\infty)$ and let us choose the boundary condition at $r=0$ to be $\psi(t,0)=0$. Such conditions correspond, for instance, to a 3-dimensional wave equation with a spherically symmetric potential (after separation of the angular dependence and multiplication of the wave function by $r$). In the 1-dimensional problem on the half-axis the boundary at $r=0$ acts reflecting on the incoming waves. 

This potential has the nice property that it does not produce any essential singularity, and thus no cuts, in the Green's function. The reason is its fast decay at infinity, i.e. at the open end. Therefore the solution can be expanded purely into quasinormal modes, without contribution of an integral along the branch cut, what allows a better insight into the problem of completeness of the expansion. Another reason is that the asymptotic behaviour of solutions $\hphi\pm(s,x)$ for big $s$ is relatively simple.

\subsection{Analytical results}

First observation is that only one end, at $r\ra\infty$, is open in this example, while the other one at $r=0$ is reflecting. It leads to a modified free Green's function
\begin{equation}
  \hG_0(s,r,r')=\frac{1}{2s}\;e^{-s|r-r'|}-\frac{1}{2s}\;e^{-s(r+r')}.
\end{equation}
The first term is the usual one for a free wave equation and the second one describes the reflection of waves at the origin.

Following the way of constructing solutions described in the previous sections, we present here only results of the consecutive steps. The solutions of the homogeneous ODE, satisfying one of the boundary conditions each, can be expressed by means of the Bessel functions
\begin{eqnarray}
  \hphi_0(s,r) &=& J_s(q)\;N_s(q e^{-r}) - N_s(q)\;J_s(q e^{-r})\\
  \hphi_+(s,r) &=& J_s(q e^{-r}),
\end{eqnarray}
where $V_0\equiv -q^2$, i.e. the potential is negative\footnote{It has no meaning as long as there are no ``bound states'' -- normalizable solutions of the ODE.}. Their Wronskian equals

\begin{equation}
  W[\hphi_0,\hphi_+](s) = \frac{2}{\pi} J_s(q)
\end{equation}
and the Green's function is
\begin{equation}
\begin{split}
  \hG(s,r,r') = &\frac{\pi}{2}\; \frac{J_{-s}(q e^{-r_<})\; J_s(q e^{-r_>})}{\sin(\pi s)} \\
                &-  \frac{\pi}{2}\; \frac{J_{-s}(q)}{\sin(\pi s) J_s(q)}\; J_s(q e^{-r'})\; J_s(q e^{-r}).
\end{split}             
\end{equation}
It can be shown that
\begin{equation}
  \hG(s,r,r') \cong \frac{1}{2s}\;e^{-s|r-r'|}-\frac{1}{2s}\;e^{-s(r+r')} \equiv \hG_0(s,r,r')
\end{equation}
as $|s|\ra\infty$.
Quasinormal mode frequencies correspond to poles of the Green's function and these are at the values of $s$ where $J_s(q)=0$. Therefore the quasinormal modes take a simple form
\begin{equation}
  \phi_n(r)=J_{s_n}(q e^{-r}).
\end{equation}
They satisfy both boundary conditions
\begin{equation}
  \phi_n(0) = 0\quad \mathrm{and}\quad \phi_n(r)\sim e^{-s_n r}\quad\mathrm{as}\quad r\ra\infty.
\end{equation}
In this example, there exist additional poles at the points where $\sin(\pi s)=0$. The residua calculated at these points cancel in integration of $\hG(s,r,r')$, but they do not cancel in integration of $e^{st}\hG(s,r,r')$, which is to be performed for inverting the Laplace transform. They produce an additional series of modes, which shows up only at intermediate times $|r-r'|<t<r+r'$ (at earlier or later times they cancel) and contributes to the Green's function by
\begin{equation}
  \frac{1}{2} \sum_{n=-\infty}^{+\infty} J_n(q e^{-r})\; J_n(q e^{-r'})\; e^{nt}.
\end{equation}
This series converges in the given range of parameters $t,r,r'$. At $r,r'\gg 1$ it can be approximated and summed up to give a closed formula
\begin{equation}
  -\frac{1}{2} + \frac{1}{2} I_0\left(q e^{-(t+r+r')/2}\right) + \frac{1}{2} I_0\left(q e^{(t-r-r')/2}\right).
\end{equation}
Finally, the Green's function is
\begin{multline}
  G(t,r,r') = \\
    \left\{
    \begin{array}{ll}
      0, & t<|r-r'| \\
      {\displaystyle \frac{1}{2} \sum_{n=-\infty}^{+\infty} J_n(q e^{-r})\; J_n(q e^{-r'})\; e^{nt},} & |r-r'|<t<r+r' \\
      {\displaystyle \sum_{n} \alpha_n J_{s_n}(q e^{-r})\; J_{s_n}(q e^{-r'})\; e^{s_n t},} & t>r+r',
    \end{array}
    \right.    
\end{multline}         
where $\alpha_n\equiv \frac{\pi}{2} N_{s_n}(q)/\left. \frac{d}{ds} J_s(q)\right|_{s=s_n}$. The solution takes the form
\begin{equation} \label{psi-r}
\begin{split}
  \psi(t,r) &= \sum_{n} B_n(t,r) J_{s_n}(q e^{-r}) e^{s_n t}\\
            &+ \sum_{n=-\infty}^{+\infty} C_n(t,r) J_n(q e^{-r}) e^{nt} 
\end{split}  
\end{equation}
with the excitation coefficients
\begin{equation} \label{coef-B}
  B_n(t,r) = \alpha_n \int_{0}^{t-r} J_{s_n}(q e^{-r'}) [s_n f(r')+g(r')]\; dr'
\end{equation}
\begin{equation} \label{coef-C}
  C_n(t,r) = \frac{1}{2} \int_{|t-r|}^{t+r} J_n(q e^{-r'}) [n f(r')+g(r')]\; dr',
\end{equation}
where we extended the definition of the initial data by $f(r)=g(r)=0$ for $r<0$.

The formulas simplify if we assume we know the support of the initial data explicitly
\begin{equation}
  supp(f) \cup supp(g) \subset [a,b].
\end{equation}
Then
\begin{enumerate}
\item \textit{Zero:} $\psi(t,r) = 0$ in two regions, $t<a-r$ and $t<r-b$, which are causally separated from initial data.
\item \textit{Sum over the additional modes:} $\psi(t,r) = \sum C_n(t,r) J_n(q e^{-r}) e^{nt}$ for $t<r+a$, i.e. in the region, where no reflection at $r=0$ had occurred yet.\\
$C_n$ become constant ($t$ and $r$ independent) in the region described by two conditions: $r-a<t<r+a$ and $t>b-r$,
i.e. where all initial data ``can be seen'' (contained in the past light cone) and before first reflected signal arrives.
\item \textit{Sum over the quasinormal modes:} $\psi(t,r) = B_n(t,r) J_{s_n}(q e^{-r}) e^{s_n t}$ in the region $t>r+a$, i.e. when some signal is already reflected at $r=0$.\\
$B_n$ become constant ($t$ and $r$ independent) for $t>r+b$, i.e. after all initial data have reflected at $r=0$.
\end{enumerate}

\begin{figure}                
\includegraphics[width=0.8\columnwidth]{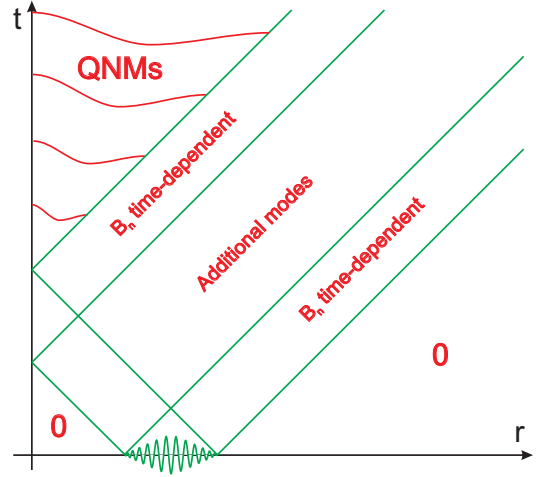}
\caption{Appearance of truncated quasinormal modes in spherical symmetry.\label{fig_r-truncated}}
\end{figure}

The reflection at the boundary $r=0$ changes the picture qualitatively (Fig. \ref{fig_r-truncated}). The quasinormal modes with constant excitation coefficients may first appear when all initial data reach the boundary and reflect. Therefore, the universal picture (sum over QNMs) shows up starting at the origin at time $t=b$ and expands with the speed of light (here $c=1$), creating a triangle (half light cone). After the time $t$, the QNM expansion is valid on $r\in[0,t-b]$.

\subsection{Numerical simulation}

We have simulated this wave equation numerically. The initial data were chosen as
\begin{equation}
  f(r) = \left\{ 
    \begin{array}{ll}
      A\cdot \left[1-\cos\left(\frac{\pi}{2} x\right)\right], & \mathrm{for}\; 0\leq r\leq 2, \\
      2\cdot A, & \mathrm{for}\; 2 \leq r \leq 22, \\
      A\cdot \left[\cos\left(\frac{\pi}{2} (x-22)\right)+1 \right], & \mathrm{for}\; 22 \leq r \leq 24,\\
      0, & \mathrm{for}\; 24 \leq r,
    \end{array}
  \right.
\end{equation}
with $A=1$ and $g(r)\equiv 0$. $f(r)$ presents a long bump, starting from $r=0$ and extending to $r=24$, smoothly connected with zero at both ends. On an interval $[2,22]$ its height is constant. The wave equation has been discretized with grid steps $dx=0.02$ and $dt=0.01$, what guarantees stability of the numerical evolution. The spatial size of the grid was $x\in[0,100]$ ($5000$ grid points), what allowed a safe evolution until $t=76$, when the first data reached the outer (free) boundary of the grid. 

The strength of the potential was chosen to be $q=2$, what corresponds to $V_0=-4$. In this case (and in general, if $V_0<0$) all poles are purely real. Numbering from right to left, i.e. from the dominant one (least damped) to more damped, their values are
\begin{equation}
  s_1 = -0.2538,\; s_2=-1.789,\; s_3=-2.961,\; s_4=-3.996, ...
\end{equation}
It can be shown analytically, that the poles of the Green's function with big $s$ (i.e. zeros of $J_s(q)$) coincide approximately with the poles of $\Gamma(s+1)$, what gives $s_n \cong -n$. Here, it agrees from $n=4$ on within $1\% 
$.

We have compared the results of numerical simulation with the solution (\ref{psi-r}). We have calculated the excitation coefficients $B_n$ and $C_n$ directly from initial data, taking into account only one, two or three least damped modes. It turned out that the differences in the damping exponents were so big, that the next modes, except the first one, were practically invisible. The reason was that when the QNM expansion began to be valid, it was $t=24$ and every mode was multiplied by $\exp(s_n t)$. These factors varied by ca. $\exp(-24)$ from one mode to another, making practically impossible to observe any mode, except the first one. 

However, it is not to be seen as a problem. We want to stress that it is rather usual that only a small number of modes contributes essentially to the sum, while all other are damped fast to zero. It is a manifestation of universality and therefore it is a highly welcome property. Not to be groundless, we studied other values of $q$, like $q=2.4$, for which $s_1=-0.003126$ and $s_2=-1.657$. The first mode was so weakly damped, that it appeared as nearly static. Then, the dynamics was well described by the sum of the first and the second mode.

\begin{figure} 
\includegraphics[width=\columnwidth]{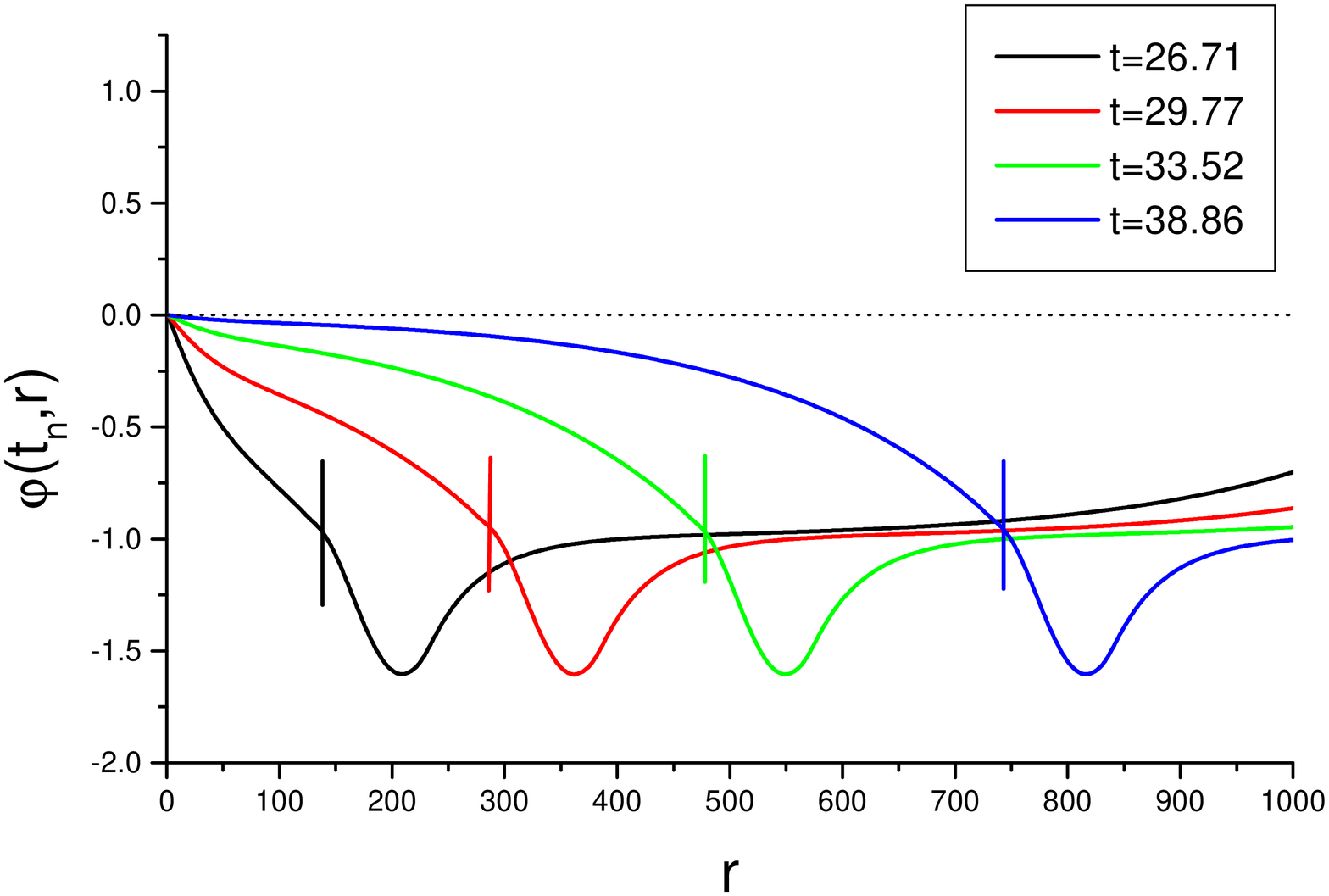}\\
\caption{\label{shots}}
\end{figure}

Coming back to the case $q=2$. Fig. \ref{shots} shows four ``screen-shots'' -- shapes of the solution at a sequence of times ($t_1=26.71, t_2=29.77, t_3=33.52, t_4=38.86$). The vertical segments show the right end of the light cone, i.e. $r=t-b$ with $b=24$, where the QNM expansion with constant coefficients is valid. Since the light cone expands in time, the segment moves to the right. In the region $r\in[0,t_n-b]$ we can see the solution approximated essentially by the first mode. Further to the right, a direct signal from the initial data, reflected at the origin, can be observed. Since it is expanded only formally into QNMs, with time and space dependent coefficients, its form is irregular and bears no sign of universality. 

\begin{figure} 
\includegraphics[width=\columnwidth]{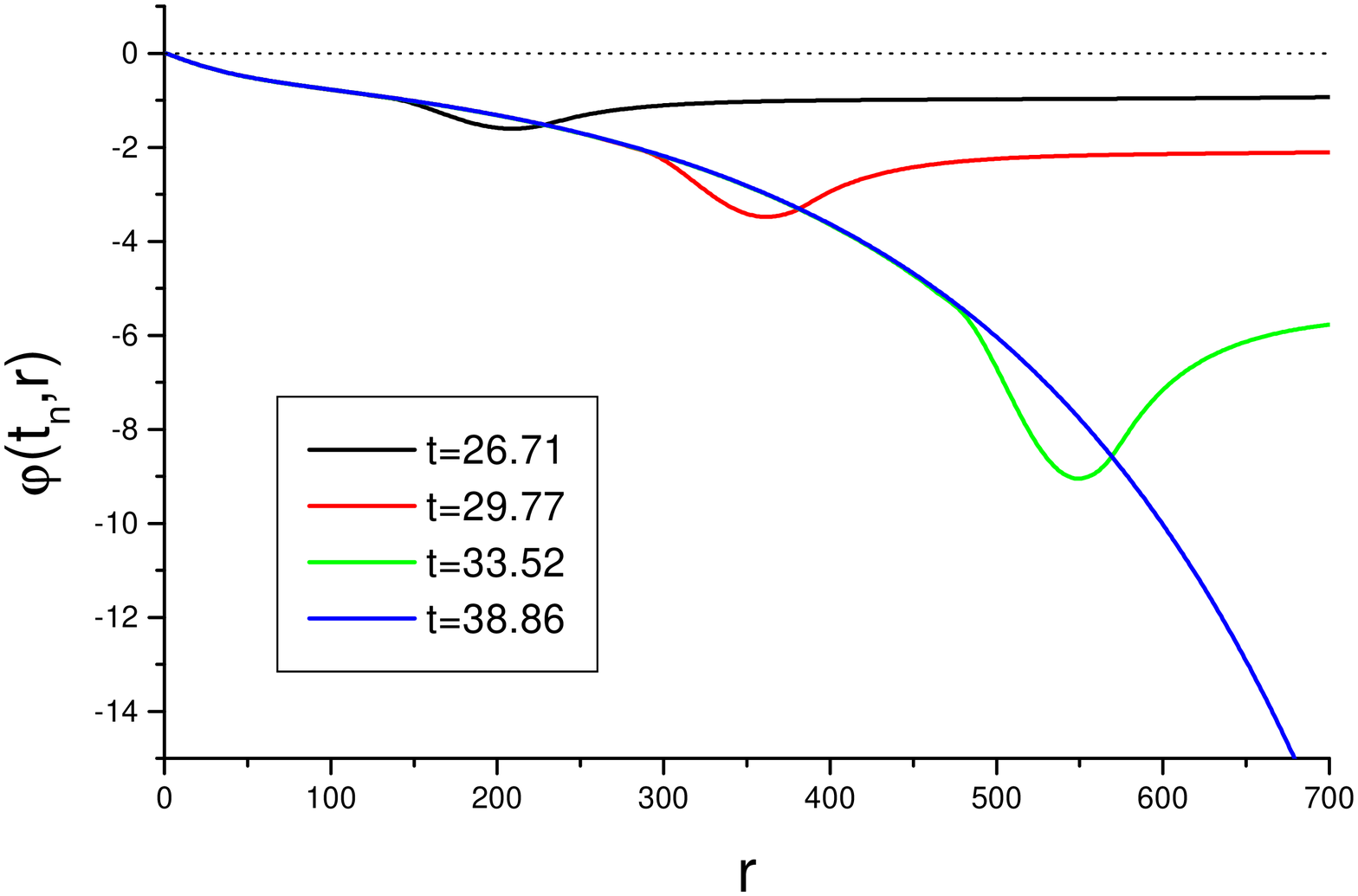}\\
\caption{\label{shots-scaled}}
\end{figure}

Since the solution, as well as the first mode, are damped in time, they appear on every ``shot'' with a different amplitude. Dividing them all by the damping factor of the first mode, $\exp(s_0 t_n)$, we can bring them to one scale, what is shown on Fig. \ref{shots-scaled}. The agreement is perfect and demonstrates that the solution is well approximated by the first mode
\begin{equation}
  \psi(t,r) \cong B_1\; J_{s_1}(q e^{-r})\; e^{s_1 t}
\end{equation}
with 
\begin{equation}
  B_1 = \alpha_1\; s_1 \int_{0}^{b} J_{s_1}(q e^{-r'})\; f(r')\; dr'.
\end{equation}


\section{Conclusions}

The main result of this paper is the time-dependent Green's function (\ref{G2}) which allows, under further restrictions on the potential $V(x)$, to express the evolution of any complactly supported Cauchy initial data by means of a sum over spatially truncated quasinormal modes. The solution for the wave function takes the form (\ref{Psi2})-(\ref{Bn2}) and is schematically shown on the figure \ref{fig_truncated}. This clarifies the appearance of QNMs in the evolution at any time, solving the problem of divergence at big distances and restoring the principle of a causal propagation. The price is to allow the excitation coefficients depend on time and space in general, but it turns out that they become constant in a special region of spacetime (roughly, common future light cone of all initial data), where the quasinormal modes become meaningful. We argue that this is the correct way to unterstand the quasinormal modes (cf. \cite{Nol-rev}) since the result is a natural consequence of solving the wave equation (\ref{waveeq}) via Laplace transform and treating the asymptotic behaviour of the integrand in the inverse transform with appropriate care, with no other ad hoc assumptions about the form of the result.

Unfortunately, the whole scheme is valid for all times ($t>0$) only for a restricted class of potentials. It seems that in general (even for fast decaying potentials) the QNMs appear later ($t>t_0(x,x')>|x-x'|$) than in an optimal case ($t>t_0(x,x')=|x-x'|$).

There remain following open problems. First, the analytic structure (singularities, cuts, asymptotics) of $\hG(s,x,x')$ in the complex plane of $s$ and its dependence on the properties of the potential $V(x)$ (asymptotics, analyticity) need further study. Second, the earliest time $t_0(x,x')$ of validitiy of the QNMs expansion and its dependence on the shape of the potential is in general unclear. They are connected through the asymptotic behaviour of $\hG(s,x,x')$ for big $s$ but the relations seem very complicated.

An interesting fact, worth further study, is the appearance of an additional series of modes in the intermediate region before the QNMs appear. However, it has been observed only in our example and it is not clear at all if this phenomenon is common.

\begin{acknowledgments}

I want to express my gratitude to Piotr Bizo{\'n} for introducing me to the issue of quasinormal modes.

\end{acknowledgments}




\appendix
\section{Normal modes, $V\equiv 0$} \label{AppA}

The homogeneous ODE (\ref{homODE}) has the following solutions
\begin{eqnarray}
  \hphi_-(s,x) & = & e^{sx}-e^{2sa} e^{-sx}\\
  \hphi_+(s,x) & = & e^{sx}-e^{2sb} e^{-sx}
\end{eqnarray}
satisfying the boundary conditions
\begin{equation}
  \hphi_-(s,a)=0,\qquad \hphi_+(s,b)=0.
\end{equation}
Their Wronskian equals
\begin{equation}
  W[\hphi_-,\hphi_+](s) = -2s \left[ e^{2sa}-e^{2sb}\right].
\end{equation}
The Laplace transformed Green's function reads
\begin{equation}
  \hG(s,x,x') = \frac{[e^{s(x_<-a)}-e^{s(a-x_<)}][e^{s(x_>-b)}-e^{s(b-x_>)}]}{2s[e^{s(a-b)}-e^{s(b-a)}]}.
\end{equation}
Its poles, the zeros of the Wronskian $W(s)$, are placed at
\begin{equation}
  s_n=\frac{n\pi i}{b-a},\qquad \mbox{for}\qquad n\in\mathbb{Z}.
\end{equation}
The explicit form of the modes
\begin{equation}
  \phi_n(x) \equiv \hphi_-(s_n,x) = \hphi_+(s_n,x)
\end{equation}
is
\begin{equation}
  \phi_n(x) = \exp\left(\frac{an\pi i}{b-a}\right) \cdot\sin\left(\frac{x-a}{b-a}n\pi\right).
\end{equation}
The time-dependent Green's function, obtained by the inverse Laplace transform, takes the explicit form
\begin{multline}
  G(t,x,x') =\\ i \sum_{n\in\mathbb{Z}} \frac{\sin\left(\frac{x-a}{b-a}n\pi\right) \cdot \sin\left(\frac{x'-a}{b-a}n\pi\right)}{n\pi} \exp\left(\frac{n\pi i}{b-a}t\right). 
\end{multline}
As we know from the alternative way of calculating the inverse Laplace transform, the series converges to zero for all $t<|x-x'|$. The series can be formally summed up and expressed as a sum of Heaviside theta functions of the parameters $t,x,x'$ and hence $\dot{G}(t,x,x')$ as a sum of Dirac delta functions -- the result of free propagation inside the interval $(a,b)$ with multi-reflections at the boundaries.

\section{Normal modes, $V\neq 0$} \label{AppB}

The homogeneous ODE, analogous to (\ref{homODE}), but with non-vanishing $V(x)$, has the following approximate solutions
\begin{multline}
 \hphi_-(s,x) =\\ \frac{1}{2}\left[
      e^{s(x-a)}\left(\alpha+\frac{\beta}{s}\right) + e^{-s(x-a)}\left(\alpha-\frac{\beta}{s}\right)\right]
      \left(1+\mathcal{O}(s^{-1})\right) \nonumber
\end{multline}       
\begin{multline}
 \hphi_+(s,x) =\\ \frac{1}{2}\left[
      e^{s(x-b)}\left(\gamma+\frac{\delta}{s}\right) + e^{-s(x-b)}\left(\gamma-\frac{\delta}{s}\right)\right]
      \left(1+\mathcal{O}(s^{-1})\right)
\end{multline}       
satisfying the Laplace transformed boundary conditions
\begin{eqnarray}
  \beta\; \hphi_-(s,a)-\alpha\; \frac{\partial \hphi_-}{\partial x}(s,a)&=&0\\
  \delta\;\hphi_+(s,b)-\gamma\; \frac{\partial \hphi_+}{\partial x}(s,b)&=&0.
\end{eqnarray}
with $\alpha, \beta,\gamma, \delta \in \R$. Their Wronskian is
\begin{multline}
  W[\hphi_-,\hphi_+](s) = \frac{s}{2} 
  \left[ e^{s(b-a)}\left(\alpha+\frac{\beta}{s}\right)\left(\gamma-\frac{\delta}{s}\right) \right. - \\
  \left.-e^{s(a-b)}\left(\alpha-\frac{\beta}{s}\right)\left(\gamma+\frac{\delta}{s}\right) \right]
  \left(1+\mathcal{O}(s^{-1})\right)       
\end{multline}
The Laplace transformed Green's function reads
\begin{multline}
  \hG(s,x,x') = \frac{1}{2s}\; \cdot \\
  \left[e^{s(x_<-a)}\left(\alpha+\frac{\beta}{s}\right) + e^{-s(x_<-a)}\left(\alpha-\frac{\beta}{s}\right)\right]\cdot \\
  \left[e^{s(x_>-b)}\left(\gamma+\frac{\delta}{s}\right) + e^{-s(x_>-b)}\left(\gamma-\frac{\delta}{s}\right)\right]\cdot \\
  \left[ e^{s(b-a)}\left(\alpha+\frac{\beta}{s}\right)\left(\gamma-\frac{\delta}{s}\right) \right. - \\
  \left.-e^{s(a-b)}\left(\alpha-\frac{\beta}{s}\right)\left(\gamma+\frac{\delta}{s}\right) \right]^{-1}\cdot \\
  \left(1+\mathcal{O}(s^{-1})\right)
\end{multline}

\section{Leaky cavity} \label{AppC}

For the leaky cavity the homogeneous ODE (\ref{homODE}) has the following solutions
\begin{equation}
 \hphi_-(s,x) = e^{sx} - e^{-sx}
\end{equation}
\begin{multline}
 \hphi_+(s,x) =\\ \frac{1}{2}\left[
      e^{s(x-L)}\left(\gamma-1\right) + e^{-s(x-L)}\left(\gamma+1\right)\right] \nonumber
\end{multline}       
satisfying the boundary conditions
\begin{eqnarray}
  \hphi_-(s,0)&=&0,\\
  s\; \hphi_+(s,L) + \gamma\; \frac{\partial \hphi_+}{\partial x}(s,L)&=&0 \label{LCbc2}
\end{eqnarray}
Their Wronskian equals
\begin{equation}
  W[\hphi_-,\hphi_+](s) = s \left[ e^{sL}(\gamma+1) + e^{-sL}(\gamma-1) \right].
\end{equation}
The Laplace transformed Green's function reads
\begin{multline}
  \hG(s,x,x') = \\
  \frac{[e^{sx_<}-e^{-sx_<}] 
        \left[e^{s(x_>-L)}\left(\gamma-1\right) + e^{-s(x_>-L)}\left(\gamma+1\right)\right]}
       {2s \left[ e^{sL}(\gamma+1) + e^{-sL}(\gamma-1) \right]}.
\end{multline}
Its poles, the zeros of the Wronskian $W(s)$, are placed at
\begin{equation}
  s_n=\frac{1}{2L}\ln\left(\frac{1-\gamma}{1+\gamma}\right) + \frac{n\pi i}{L},
 \qquad \mbox{for}\qquad n\in\mathbb{Z}.
\end{equation}
The explicit form of the modes
\begin{equation}
  \phi_n(x) \equiv \hphi_-(s_n,x) = \frac{(-1)^n}{\sqrt{1-\gamma^2}}\; \hphi_+(s_n,x) 
\end{equation}
is
\begin{multline}
  \phi_n(x) = 2\;\sinh(s_n x) = \\ 
  \exp\left(\frac{n\pi i x}{L}\right) \sqrt{\frac{1-\gamma}{1+\gamma}}^{(x/L)} -
  \exp\left(\frac{-n\pi i x}{L}\right) \sqrt{\frac{1+\gamma}{1-\gamma}}^{(x/L)}.
\end{multline}
The time-dependent Green's function, obtained by the inverse Laplace transform, takes the form
\begin{multline}
  G(t,x,x') = \sum_{n\in\mathbb{Z}} 
  \frac{\sinh(s_n x) \cdot \sinh(s_n x')}{s_n} \exp(s_n t). 
\end{multline}
As we again know from an alternative way of calculating the inverse Laplace transform, the series converges to zero for all $t<|x-x'|$.

\bibliography{QNMs}
\bibliographystyle{unsrt}

\end{document}